\documentclass[paper,onecolumn,long,noendfloat,new,nonreprod,noblind]{geophysics}

\usepackage{float} 
\usepackage{graphicx} 
\usepackage{caption} 
\usepackage{mathtools}
\usepackage{bm}
\usepackage{amsmath}
\usepackage{amsfonts}
\usepackage{amssymb}
\usepackage{multirow}
\usepackage{placeins}
\usepackage{layouts}
\usepackage[font=small,labelfont=small]{caption}
\usepackage[shortcuts]{extdash}
\newcommand{\dr}{\partial}

\everymath{\displaystyle}

\renewcommand{\figdir}{Fig}

\begin{document}

\title{Moment tensor inversion of perforation shots using distributed acoustic sensing}

	

 \author{Milad Bader\footnotemark[1], Robert G. Clapp\footnotemark[1], Kurt T. Nihei\footnotemark[2], and Biondo Biondi\footnotemark[1]}

    \righthead{Perforation MT inversion using DAS}


	\footnotetext[1]{Stanford University, Department of Geophysics, 397 Panama Mall, Stanford, CA 94305.\\E-mail: nmbader@sep.stanford.edu; bob@sep.stanford.edu; biondo@sep.stanford.edu.}
    \footnotetext[2]{Chevron Technology Center, 1500 Louisiana Street, Houston, TX 77002.\\E-mail: kurttnihei@chevron.com.}


\maketitle

\begin{abstract}
Distributed acoustic sensing (DAS) fibers have enabled various geophysical applications in unconventional reservoirs. Combined with perforation shots, a DAS fiber can record valuable guided waves that propagate in the reservoir formation and carry information about its properties. However, the representation of perforation shots as seismic sources, needed to conduct quantitative analysis, remains unknown. We model such sources using a superposition of three mechanisms for which we derive the moment tensor representation. Using field DAS data recorded in the same well where the perforations are located, we establish a workflow to invert the resolvable components of the total moment tensor for 100 different perforation shots. By scrutinizing the inversion results, we conjecture that the moment tensor can indicate how effectively a perforation shot creates micro-cracks in the surrounding rock. Furthermore, our inverted moment tensors form the basis for a subsequent elastic full-waveform inversion.
\end{abstract}
\section{Introduction}

Distributed acoustic sensing (DAS) has become an established technology that allows recording the average over the gauge length of the seismic strain (or strain rate) wavefield with an unprecedented temporal and spatial resolution for various geophysical applications \cite[]{li2022distributed}. In particular, downhole DAS has been used in vertical seismic profiling (VSP) for seismic imaging and time-lapse reservoir monitoring \cite[]{mateeva2014distributed}, acoustic impedance inversion \cite[]{kazei2021inverting}, and elastic full-waveform inversion \cite[]{egorov2018elastic,eaid2020multiparameter}.

In unconventional reservoirs, a DAS fiber deployed in the horizontal section of the well has enabled a range of new geophysical applications. \cite{huot2022detection} used supervised machine learning to automatically detect microseismic events recorded by DAS in a horizontal well. \cite{eaid2021estimation} leveraged a deep neural network architecture to estimate the microseismic source mechanisms from DAS data. \cite{luo2021near} observed rare near-field strain signals from microseismic events on DAS records. \cite{lellouch2022microseismic} used guided waves recorded by a single DAS fiber to locate microseismic events induced by hydraulic stimulation. \cite{luo2021seismic} leveraged similar microseismic-induced guided waves to conduct dispersion curves inversion and derive reservoir elastic properties. \cite{jin2017hydraulic} utilized the low-frequency DAS recording (below 0.5 Hz) to monitor hydraulic fracture generation and propagation. \cite{lellouch2020fracture} estimated geometrical fracture properties from high-frequency elastic waves (50 to 700 Hz) generated by perforation shots and recorded by a DAS fiber. \cite{schumann2022inferring} characterized fracture-well connectivity by measuring the decay rate of tube waves generated by perforation shots.

In this work, we leverage the elastic waves generated by the perforating gun and recorded by an \textit{in situ} DAS fiber to characterize the seismic source representing the action of that gun on the surrounding rock formation. Our aim is to provide an accurate representation of the source mechanism to be used as a perforation indicator and in subsequent full-waveform inversion for reservoir properties estimation. To our knowledge, no source model representing a perforation shot has been proposed or estimated in the literature. \cite{fehler1984cross} modeled an explosive source in a fluid-filled borehole using a diagonal moment tensor (MT). We represent a perforation shot by a superposition of three source mechanisms and derive the corresponding MTs for an isotropic and anisotropic VTI media. We determine the resolvable components of the resulting total MT given the field DAS data geometry and the medium elastic model. We set up a processing workflow to estimate these components for 100 perforation shots from 20 hydraulic stimulation stages. By scrutinizing the inversion results, we show that the MTs may give insights into the rock fracturing caused by the perforations. Moreover, these MTs serve as a basis to conduct subsequent elastic full-waveform inversion to estimate the reservoir elastic properties.


\section{Field data description}

The data treated in this work is acquired by a DAS fiber cemented behind the steel casing of a deviated monitor well drilled across an unconventional reservoir formation. The quasi horizontal reservoir layer is about 15 m thick and has been hydraulically stimulated following two main steps. The first step consists of perforating the well casing and surrounding rock using a gun designed for that purpose. In the second step, a high-pressure fluid is injected through the newly created perforations to create fractures in the reservoir rock and increase its hydraulic permeability. The horizontal section of the well, aligned in the $x$-direction, is subdivided into stages of approximately 50 m length and each stage is perforated at five or six locations approximately 10 m apart before the fluid injection step. A plug is used to hydraulically isolate a current stage while being fractured from the previously stimulated one. Figure \ref{fig:sketch_perforation} illustrates the perforation procedure. Each perforation shot consists of a cluster of several explosive-like charges. The charges are directed in the plane orthogonal to the well ($y-z$ plane) but their phasing angle in that plane is unknown. Note that different charges within the same perforation cluster may have different phasing angles, but all charges must be oriented in a way to avoid damaging the DAS fiber.

\plot[!h]{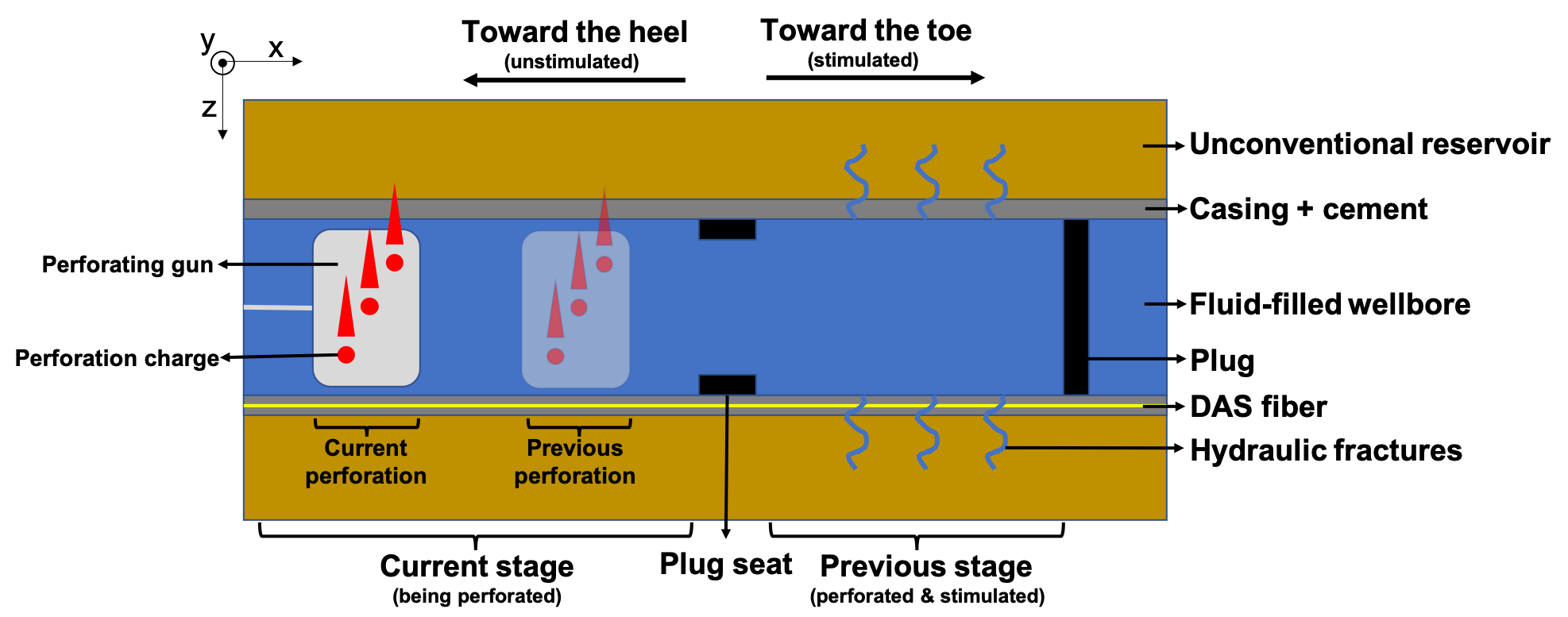}{width=0.95\textwidth}{Sketch illustrating the perforation procedure in a horizontal well.}

The DAS fiber records the seismic wavefield generated by the perforating gun acting as a seismic source. In this work, the fiber measures the average strain rate along a 10 m gauge length with temporal and spatial samplings of 0.5 ms and 1 m, respectively. Figure \ref{fig:raw_das_example} shows an annotated DAS shot gather example. The unconventional reservoir layer is a waveguide that traps P- and S-waves over hundreds of meters as described in \cite{lellouch2019observations}. Strong tube waves corresponding to acoustic waves traveling inside the fluid-filled wellbore can also be observed, although we treat them in this work as noise. We utilize part of P- and S-arrivals recorded at negative offsets to estimate the source parameters. These arrivals have traveled in the unstimulated part of the reservoir layer. Hence, they are not affected by the (unknown) perturbations caused by
hydraulic stimulation and yield more robust and accurate source inversion results. Note also the strong horizontal event at about 0.02 sec caused by the extreme deformation of the DAS fiber in proximity of the perforating gun.

\plot[!h]{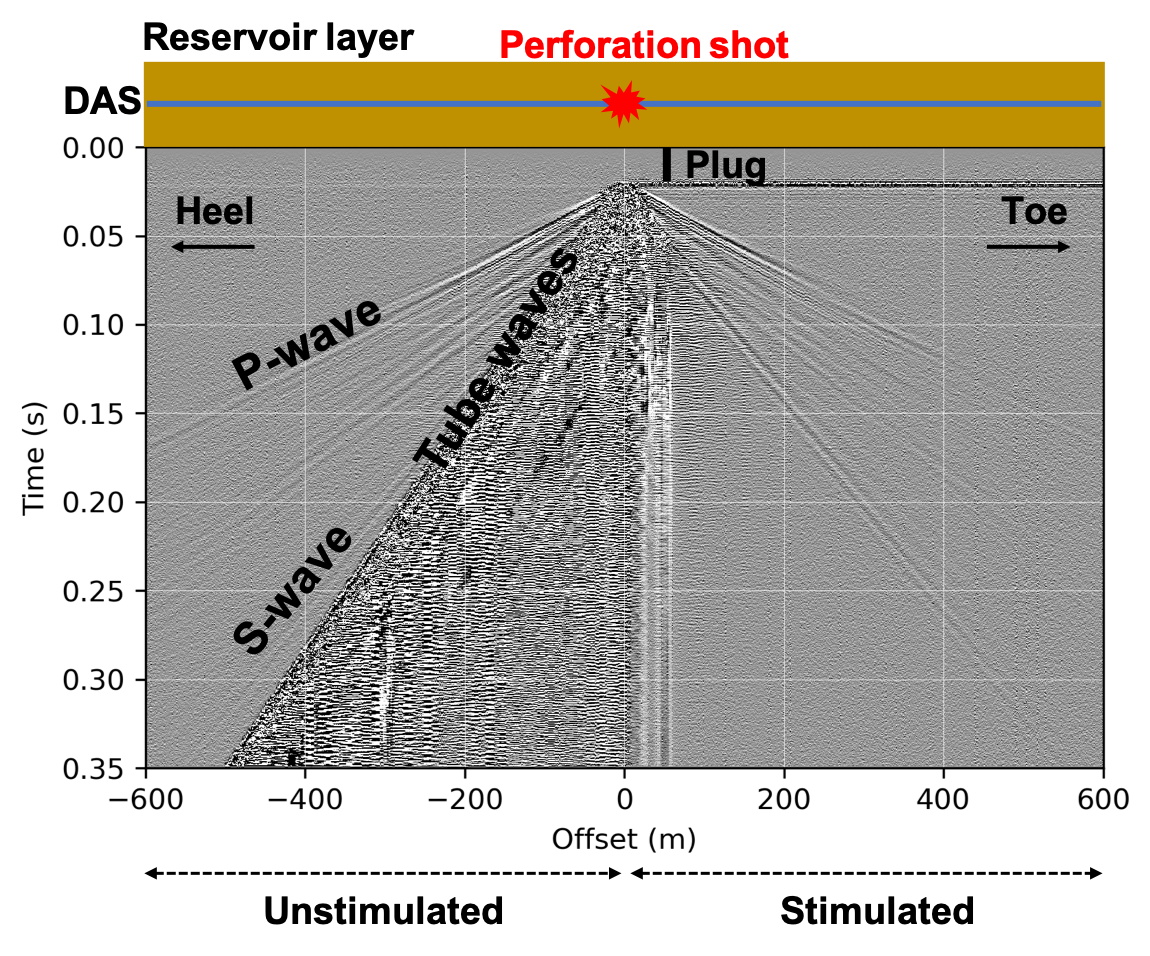}{width=0.95\textwidth}{Raw DAS shot gather example with a sketch on top depicting the acquisition geometry. Reservoir-guided P- and S-waves are recorded by the DAS fiber, along with strong tube waves traveling inside the fluid-filled wellbore. We use a part of P- and S-waves recorded at negative offsets to conduct our moment tensor inversions.}

\section*{Method}

\subsection*{Moment tensor representation}
A perforation charge acts as a seismic source and generates seismic waves that propagate either inside the fluid-filled borehole as tube waves \cite[]{schumann2022inferring} or in the surrounding rock formations as body or guided waves \cite[]{lellouch2019observations}. Considering only body waves far from the perforation charge, we can represent the latter with a point-source time-dependent moment tensor
\begin{equation}\label{mt}
\mathbf{M}_t = \mathbf{M} w(t),
\end{equation}
where $\mathbf{M}$ is a time-independent symmetric $3\times 3$ tensor, and $w$ is a normalized pulse (source time function). For a review of moment tensors and their representation of seismic sources we refer the reader to \cite{jost1989student}. To derive the structure of $\mathbf{M}$, we decompose the action of a perforation charge on the surrounding rock into the three elementary source mechanisms taking place and illustrated in Figure \ref{fig:sketch_mechanism}:
\begin{itemize}
    \item A cylindrical explosion (ce) inside the fluid-filled borehole which is manifested by a change in the well cross-section area (red arrows in Figure \ref{fig:sketch_mechanism}) over a characteristic length along the well \cite[]{fehler1984cross}.
    \item A dipole force (df) that is due to the direct mechanical effect of the perforation charge on the well wall in the direction given by the phasing angle $\theta$. In fact, a given charge exerts a single directional force on the cylindrical wall but this force is balanced by an effective force in the opposite direction (conservation of momentum).
    \item A cylindrical opening (co) that occurs when the charge travels through the formation, creating a cylindrical hole which expands radially as indicated by the red arrows in Figure \ref{fig:sketch_mechanism}.
\end{itemize}
Then, we write $\mathbf{M}$ as a linear superposition of three tensors
\begin{equation}\label{mt-decomposed}
    \mathbf{M}=\Big(\mathbf{M}^{ce}+\mathbf{M}^{df}+\mathbf{M}^{co}\Big),
\end{equation}
where $\mathbf{M}^{\dots}$ are the time-independent MTs corresponding to the three source mechanisms described above. For a later comparison, we also show in Figure \ref{fig:sketch_mechanism} the tensile crack (tc) mechanism \cite[]{aki2002quantitative} corresponding to a planar crack with a dislocation along the well ($x$-direction).

\plot[!h]{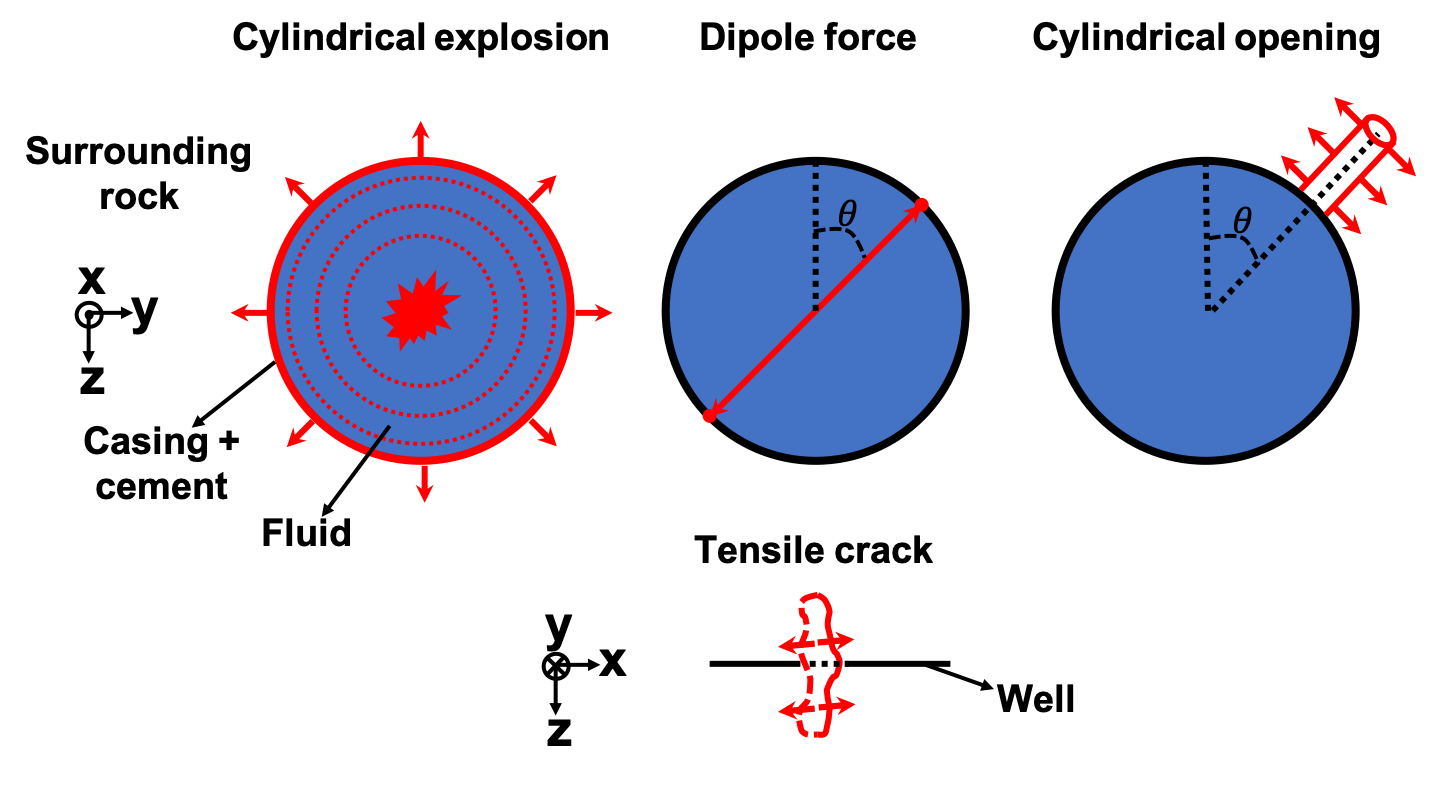}{width=0.95\textwidth}{(Top) Sketch illustrating the well cross-section with the three seismic source mechanisms corresponding to a perforation charge: cylindrical explosion (ce), dipole force (df), and cylindrical opening (co). (Bottom) Illustration of the tensile crack (tc) mechanism. The red arrows indicate the dislocation caused by each mechanism.}

We derived the MT structure of each elementary mechanism using the general form \cite[]{aki2002quantitative}
\begin{equation}\label{mt-general}
M_{kl}=\iint_S[u_{i}]\nu_j c_{ijkl}\,dS, \quad \text{summation over $i,j$},
\end{equation}
where $S$ is the surface of dislocation at the source location, $[u_i]$ is the displacement jump over the surface element $dS$ in the $i^{th}$-direction, $\nu_j$ is the $j^{th}$-component of the unit vector normal to $dS$, and $c_{ijkl}$ are the elements of the stiffness tensor of the rock surrounding the well at the source location. Note that the $1^{st}$, $2^{nd}$, and $3^{rd}$ directions correspond to the Cartesian $x$-, $y$-, and $z$-directions, respectively. The structures of the elementary MTs read
\begin{equation}\label{mt-elementary}
    \mathbf{M}^{ce} = 
    \begin{pmatrix}
        \times & 0 & 0 \\
        0 & \times & 0\\
        0 & 0 & \times
    \end{pmatrix}, \quad 
    \mathbf{M}^{df} =
    \begin{pmatrix}
     0 & 0 & 0 \\
     0 & \times & \times \\
     0 & \times & \times
    \end{pmatrix}, \quad
    \mathbf{M}^{co} =
    \begin{pmatrix}
    \times & 0 & 0 \\
     0 & \times & \times \\
     0 & \times & \times
    \end{pmatrix},
\end{equation}
where the $\times$ correspond to potentially non-zero values. We give in Appendix \ref{appendix:mt} the explicit expressions of the above MTs for surrounding isotropic and anisotropic VTI media. Note that $\mathbf{M}^{df}$ and $\mathbf{M}^{co}$ depend on the phasing angle $\theta$ whereas $\mathbf{M}^{ce}$ does not. The structure of $\mathbf{M}^{ce}$ in an isotropic medium was previously used to represent an acoustic source in a fluid-filled borehole \cite[]{fehler1984cross} as well as an acoustic-resonator in a magma-filled volcanic pipe \cite[]{chouet1985excitation}. It follows from expressions \eqref{mt-decomposed} and \eqref{mt-elementary} that the total MT has the structure
\begin{equation}\label{mt-total}
    \mathbf{M} = 
    \begin{pmatrix}
        \times & 0 & 0 \\
        0 & \times & \times \\
        0 & \times & \times
    \end{pmatrix},
\end{equation}
with at most four independent non-zero components. The scalar seismic moment of $\mathbf{M}$, denoted $M_0$, is determined according to \cite[]{silver1982optimal}
\begin{equation}\label{seismic-moment}
    M_0 = \frac{1}{\sqrt{2}} \Vert \mathbf{M} \Vert_F,
\end{equation}
where $\Vert \cdot \Vert_F$ designates the Frobenius norm.

\subsection*{Forward and inverse problems}  

The seismic displacement field $\mathbf{u}$ in the $n^{th}$ direction is related to $\mathbf{M}$ via the relation \cite[]{aki2002quantitative}
\begin{equation}\label{displacement-expression}
    u_n = M_{ij} w * \dr_j G_{ni}, \quad \text{(summation over $i,j$)},
\end{equation}
where $G_{ni}$ are the elements of Green's tensor, and $*$ denotes time convolution. For a DAS channel located at $\mathbf{x}=(x,y,z)^T$ and measuring the average strain over a gauge length $g$ along the fiber, the modeled data can be written as
\begin{equation}\label{das-general}
    d(t,\mathbf{x}) = \frac{1}{g}\int_{-g/2}^{g/2} \mathbf{l}^T(\mathbf{x}(s)) \, \boldsymbol{\epsilon} (t,\mathbf{x}(s)) \, \mathbf{l}(\mathbf{x}(s))\,ds,
\end{equation}
where $\boldsymbol{\epsilon} = (\boldsymbol{\nabla} \mathbf{u} + \boldsymbol{\nabla} \mathbf{u}^T)/2$ is the symmetric strain tensor, $\mathbf{l}$ is the directional unit vector tangent to the fiber, and $s$ is the curvilinear coordinate along the fiber relative to the channel location. Note that the general expression \eqref{das-general} is valid for a DAS fiber with arbitrary shape and is equivalent to the expression given by \cite{eaid2020multiparameter} in terms of Frenet-Serret coordinate system (local system describing the fiber). When the DAS fiber is approximately straight over the gauge length (e.g. the deviation from a straight line is small compared to the seismic wavelength of interest), the general expression \eqref{das-general} simplifies to
\begin{equation}\label{das-linear}
    d(t,\mathbf{x}) = \frac{1}{g}\Big( \mathbf{u}(t,\mathbf{x}+\frac{g}{2}\mathbf{l}(\mathbf{x})) - \mathbf{u}(t, \mathbf{x}-\frac{g}{2}\mathbf{l}(\mathbf{x})) \Big)^T \mathbf{l}(\mathbf{x}).
\end{equation}
Thus, modeling a single DAS channel amounts to extracting two point measurements from the displacement field $\mathbf{u}$, taking their difference, and projecting onto the fiber direction at the channel location $\mathbf{x}$. For a strain rate modeling, it suffices to take the time derivative of $d$ or replacing $w$ by its time derivative $\dot{w}$.

Combining equations \eqref{displacement-expression} and \eqref{das-linear} for multiple DAS channels, we write the linear forward problem in the compact form
\begin{equation}\label{forward-problem}
    \mathbf{d} = \mathbf{A} \mathbf{m},
\end{equation}
where $\mathbf{m}=(M_{11},M_{22},M_{33},M_{23})^T$ is the flattened MT, and $\mathbf{A}$ is a matrix (modeling operator) that we construct numerically column by column. Each column corresponds to a set of synthetic DAS data modeled using a canonical MT $\mathbf{m}=\mathbf{e}_j$ (vector of all zeros except at the $j^{th}$ location). Then, we estimate $\mathbf{M}$ according to
\begin{equation}\label{inverse-problem}
    \hat{\mathbf{m}} = \mathbf{A}^{\dagger} \mathbf{d}_{obs},
\end{equation}
where $\mathbf{A}^{\dagger}$ is the pseudo-inverse of $\mathbf{A}$, and $\mathbf{d}_{obs}$ is the observed (field) DAS data. Thus, $\hat{\mathbf{m}}$ gives the minimum length least-squares solution for $\mathbf{M}$ \cite[]{aster2018parameter}.

\subsection*{Radiation patterns and resolvability}

We show in Figure \ref{fig:radiation_pattern} the 3D far-field radiation patterns in the $u_1$ displacement wavefield for the individual components of $\mathbf{m}$, assuming a point source located at the origin $\mathbf{x}=(0,0,0)^T$. Since the DAS fiber is mostly aligned with the source in the $x$-direction ($y=z=0$), it is only sensitive to $u_1(t,\mathbf{x}=(x,0,0)^T)$ according to expression \eqref{das-linear}. It follows that, in an isotropic and homogeneous medium, DAS can only record P-waves generated by the $M_{11}$ component of the seismic source. However, due to the medium heterogeneity, in particular the waveguide character of the reservoir layer, DAS can record any $u_1$ wavefield radiating in the $x-z$ plane. Therefore, DAS is sensitive to P- and S-waves generated by $M_{11}$ and $M_{33}$ components. This is confirmed by the observation of S-waves in the field shot gather shown in Figure \ref{fig:raw_das_example}. The other MT components $M_{22}$ and $M_{23}$ are not resolvable by the inversion \eqref{inverse-problem} and we can reduce the space of unknowns to two dimensions where the reduced model becomes $\mathbf{m}=(M_{11},M_{33})^T$. Moreover, we define the observable seismic moment as
\begin{equation}\label{seismic-moment-observable}
    M_0^{obs} = \sqrt{M_{11}^2 + M_{33}^2}\,.
\end{equation}

\plot[!ht]{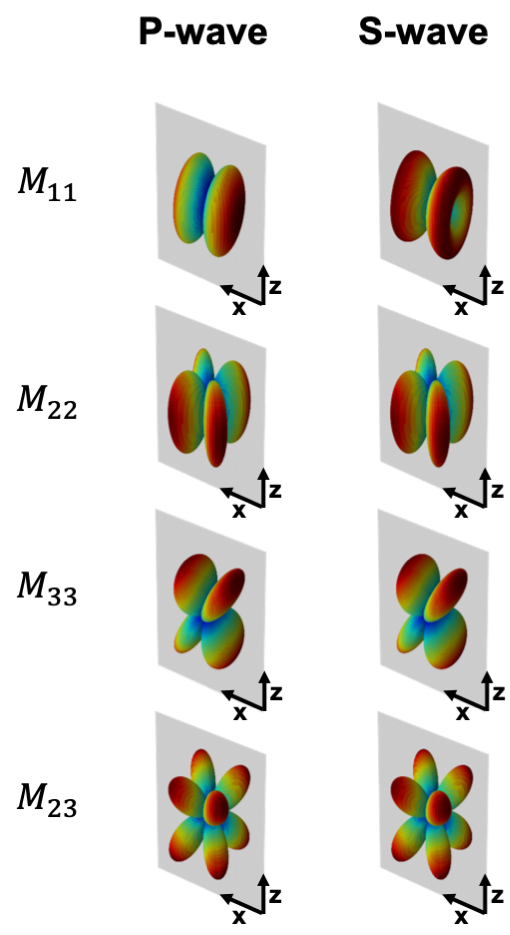}{width=0.7\textwidth}{Horizontal displacement ($x$-direction) far-field 3D radiation patterns for the non-zero components of the total MT in an isotropic homogeneous medium. The $y$-direction (not shown) is orthogonal to the $x-z$ plane. Warm colors indicate large displacement magnitude. A far-field DAS fiber installed in the x-direction would only record the P-wave generated by the $M_{11}$ component.}
\section*{Workflow and results}

We start by building the elastic VTI model needed to construct the columns of the matrix $\mathbf{A}$. We derived vertical $V_{P0}$, $V_{S0}$, and density profiles using nearby vertical well logs. Then, we constructed the anisotropy parameter $\epsilon$ \cite[]{thomsen1986weak} by calibrating the normalized Gamma-ray using the seismic data. Given the acquisition geometry, we cannot constrain the $\delta$ model accurately. Therefore, we assumed $\delta=\frac{1}{2}\epsilon$. We extrapolated all profiles laterally in the $x$-direction using a horizon conformal to the reservoir formation which has a constant thickness. The geology is quite stable and practically flat in the lateral $y$-direction. Thus, we assume a homogeneous elastic model along that direction and perform our synthetic DAS modeling in 2D (in the $x-z$ plane). From the continuous DAS recordings we were able to extract shot gathers for 100 perforation sources (numbered 0 to 99) in the monitor well taken from 20 different stages (not all consecutive). Figure \ref{fig:vp} shows the location of the shots and the well trajectory including the DAS fiber along with the $V_{P0}$ model. The perforation procedure described in Figure \ref{fig:sketch_perforation} is carried out from the rightmost side of the well (source number 0) to the leftmost side (source number 99). Given that both the source pulse and the MT are unknown, we estimated them separately following a four-step workflow:

\plot[!h]{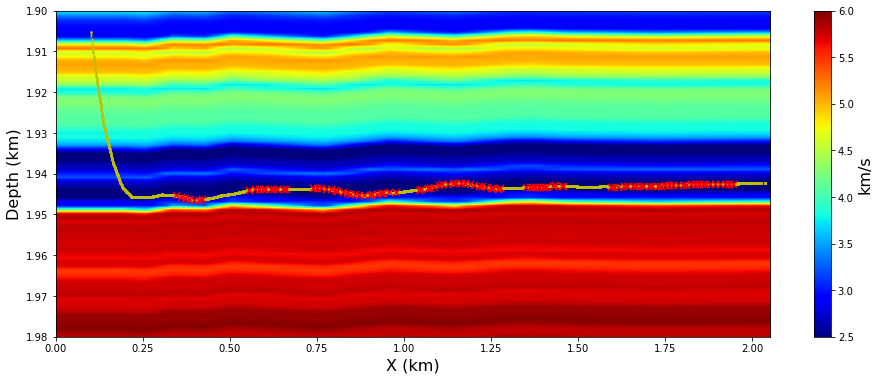}{width=0.95\textwidth}{Vertical P-wave velocity model ($V_{P0}$) part of the 2D VTI model used to generate synthetic DAS data. The yellow line indicates the well trajectory (and DAS fiber) and the red crosses correspond to 100 perforation shots in the low-velocity unconventional reservoir layer.}

\subsection{Step 1: Data processing}

We start by windowing the DAS shot gathers to remove tube waves and keep a limited offset range of [$-450$ m, $-200$ m]. Then, we apply a wavelet correction filter \cite[]{auer2013critical} to compensate for 3D-to-2D effects. We also band-pass the data in the range [$30$ Hz, $100$ Hz] and apply mild f-k filtering to remove steep noise in the x-t domain. Finally, we subsample in time from 0.5 to 2 ms to reduce the data size and speedup the inversion \eqref{inverse-problem}.

\subsection{Step 2: Source pulse calibration}

As the source pulse is unknown \textit{a priori}, we set up to estimate it from the data. To ensure robust MT inversion, we adopt an analytical time function of the form
\begin{equation}\label{wavelet}
    w(t)=\frac{a}{\Big(e^{(t_0-t)/\sigma_1}+e^{(t-t_0)/\sigma_2}\Big)^2}.
\end{equation}
that we calibrate using a limited number of degrees of freedom. This time function is a generalization of Hubbert's Peak \cite[]{hubbert1956nuclear} that allows different rise and decay times $\sigma_1$ and $\sigma_2$ (J. F. Claerbout and S. Ronen, personal communication, 2022), compared to a Gaussian or an exponential pulse. The normalization factor $a$ is determined such as $\int_{-\infty}^{+\infty}|\dot{w}(t)|dt=1$. We also allow a time shift through the parameter $t_0$ since the origin time of the perforation shots is not accurately known.


For each perforation shot, we performed a grid search to determine the optimal triplet $(\sigma_1, \sigma_2,t_0)$ such that $\sigma_2 \leq \sigma_1$ (decay time longer than rise time). For each triplet, we built the corresponding matrix $\mathbf{A}$ and inverted the MT \eqref{inverse-problem}. The optimal triplet corresponds to the smallest residual norm $\Vert \mathbf{A}\hat{\mathbf{m}}-\mathbf{d}_{obs}\Vert_2$. At this point, the inverted MTs are only provisional and serve as a proxy to determine the optimal source pulse and generate a set of synthetic data comparable to field data. Figure \ref{fig:func_global} shows the residual norm corresponding to the grid search procedure for shot \#49. Warm (cool) color corresponds to large (small) residual and the white dot corresponds to the optimal triplet. We show in Figure \ref{fig:das_obs_syn} the processed field data from Step 1 and the modeled DAS data corresponding to the optimal source pulse for the same shot \#49.

\plot[!h]{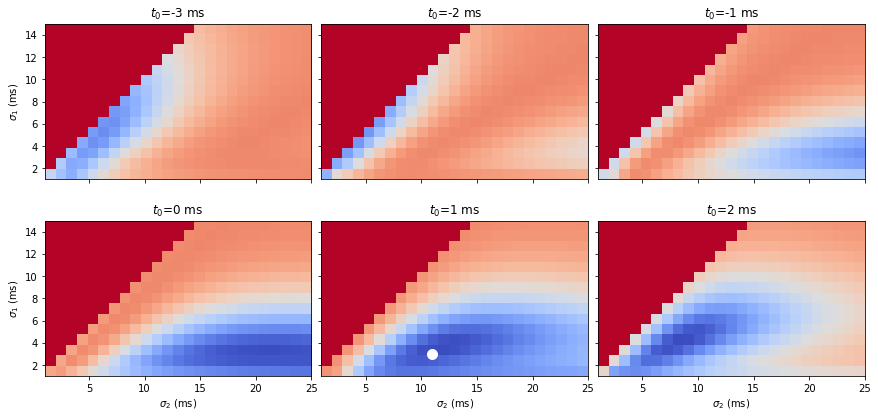}{width=0.95\textwidth}{Residual norm $\Vert \mathbf{A}\hat{\mathbf{m}}-\mathbf{d}_{obs}\Vert_2$ for the grid search procedure to determine the source pulse parameters $\sigma_1$, $\sigma_2$, and $t_0$. Warm (cool) color corresponds to high (low) values. The white dot indicates the optimal triplet.}

\plot[!h]{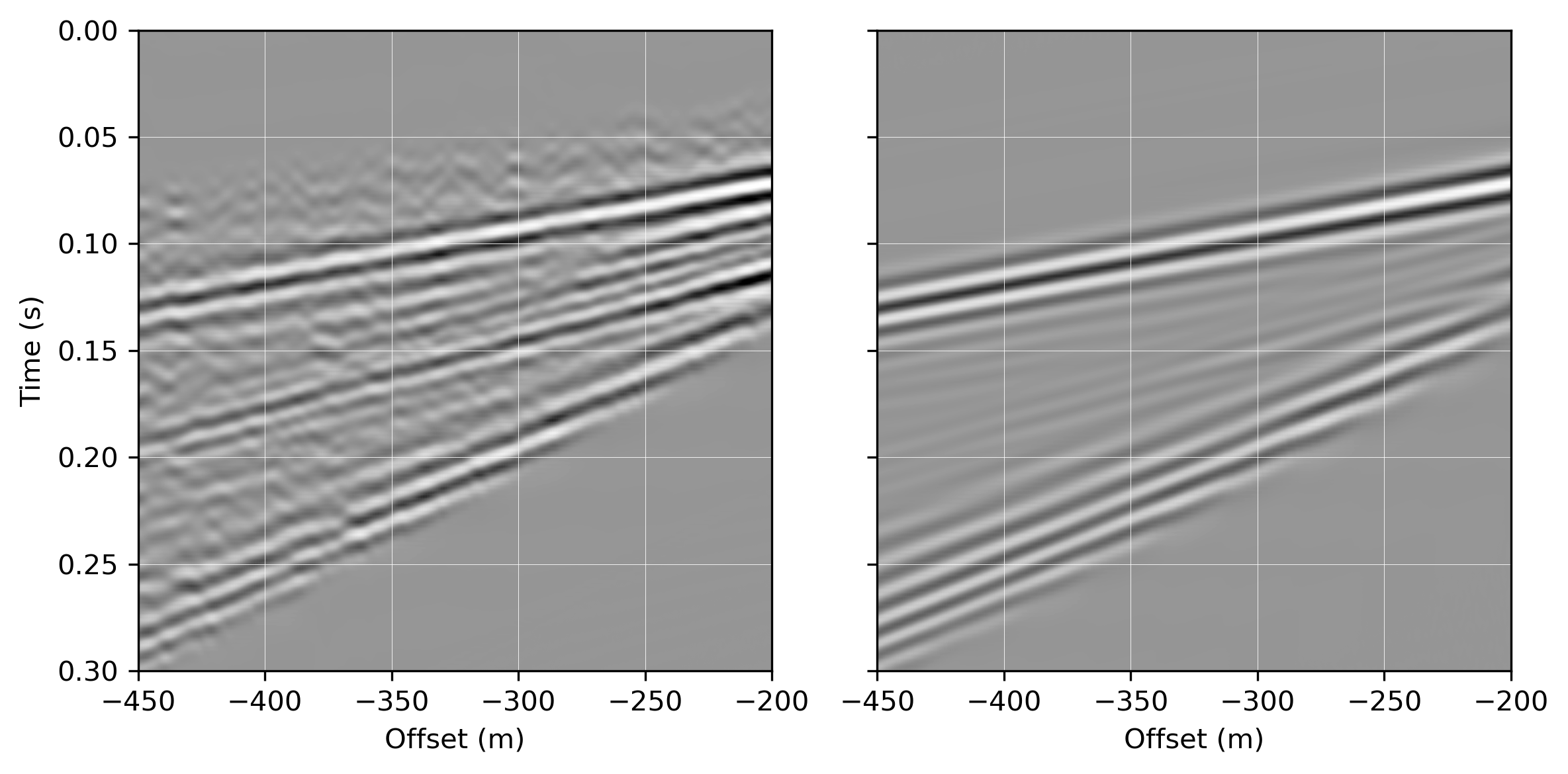}{width=0.95\textwidth}{(Left) Processed Field DAS data from Step 1, and (Right) modeled data from Step 2 for the perforation shot \#49. Both images have the same amplitude scale.}

\subsection{Step 3: Data calibration}

To ensure robust MT inversion, we calibrated the field data to match the synthetic data (Figure \ref{fig:das_obs_syn}) in terms of amplitude decay with offset and time alignment. This calibration corrects, at least partially, for the 3D-to-2D amplitude discrepancies, the absence of attenuation in our DAS modeling, and the potential inaccuracies in our elastic model. To derive the amplitude correction function, we estimated the average (over shots) amplitude decay with offset using the peak signal envelop for both field and modeled data. Then we performed a linear fitting as shown in Figure \ref{fig:amplitude_decay}. We deduced the amplitude correction function to be applied to every field shot gather
\begin{equation}\label{amplitude-correction}
    f(x)=\Big(\frac{x}{x_0}\Big)^{q-p},
\end{equation}
where $x=\log_{10}(\text{offset})$, $x_0=\log_{10}(200)$, $p$ (respectively $q$) is the estimated amplitude decay rate for field (respectively modeled) data, and 200 m is the minimum offset we use in the analysis. Here $q-p=0.27$. Note that we only calibrate the amplitude decay rate and not the absolute amplitudes because the latter is accounted for in the final MT inversion.

\plot[!h]{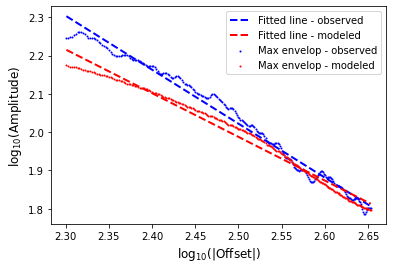}{width=0.95\textwidth}{Average (over shots) amplitude decay with offset for field and modeled data. We built the curves using the peak envelop, then we fitted lines to each curve to derive a correction function.}

Next, we performed a trace-by-trace alignment of field data with modeled data based on the maximum cross-correlation. Then, we computed the normalized cross-correlation coefficients (NCC) between the two sets of traces and dropped 35\% of them for each shot gather based on the 35 percentile NCC. The trace drop aims at removing bad and low signal-to-noise (SNR) traces before final MT inversion.

\subsection{Step 4: Final MT inversion}

Given the data quality disparity from shot to shot, and assuming that the source pulse is the same for all shots, we selected the median values (over shots) of the time function parameters from Step 2. Then, we re-built the matrix $\mathbf{A}$ for each shot while accounting for the trace drop determined in Step 3, and inverted the calibrated data according to \eqref{inverse-problem}. We show in Figure \ref{fig:condition_number} the histogram of the condition number of $\mathbf{A}$ for all shots. The low values of this number (order $O(1)$) is another confirmation that both MT components $M_{11}$ and $M_{33}$ are well resolved from DAS data even though the DAS fiber is aligned with the sources in the $x$-direction. 

\plot[!h]{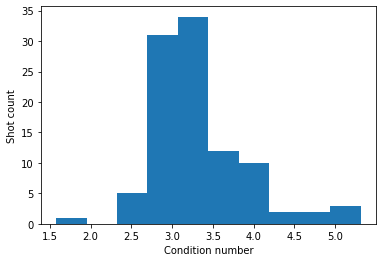}{width=0.95\textwidth}{Histogram of condition number of the forward modeling matrix $\mathbf{A}$. All condition numbers are of the order $O(1)$ which confirms the proper resolvability of both MT components $M_{11}$ and $M_{33}$.}

\subsection{Results and discussion}

We cross-plot the final inverted $M_{11}$ and $M_{33}$ and their respective histograms in Figure \ref{fig:mt_scatter_histogram}. Given the uncertainty in relating the absolute amplitudes of our DAS data to the actual strain wavefield expressed in m/m, we give the MT components a scaled N.m unit with the scaling factor being unknown. This scaling factor depends on the conversion between DAS measurement and strain, and is identical for all shots. The scatter plot is colored by the observable seismic moment \eqref{seismic-moment-observable} and dark (bright) colors correspond to larger (smaller) values. We observe two main clusters of perforation shots. The first one is characterized by small observable seismic moment and limited variability in MT components. The second has large observable seismic moment driven mainly by larger $M_{11}$ component. Note also the Gaussian-like distribution of $M_{33}$ and the exponential-like distribution of $M_{11}$. The existence of two clusters and the distribution discrepancy of the two MT components indicate changes in the seismic sources that cannot be attributed to random processes.

To further investigate these changes, we plot in Figure \ref{fig:mt_ratio} the ratio $M_{33}/M_{11}$ for the inverted MTs and for the four elementary source mechanisms illustrated in Figure \ref{fig:sketch_mechanism} (refer to Appendix \ref{appendix:mt} for the MT expression of these mechanisms). This plot allows us to distinguish between high $M_{33}/M_{11}$ ratio sources dominated by cylindrical explosion mechanism from low-ratio sources dominated by crack-opening mechanisms (cylindrical opening and tensile crack). For sources with low ratios, the $M_{11}$ component, which is proportional to displacement jump in the $x$-direction (see expression \eqref{mt-general}), can be a semi-quantitative measure of the crack width generated by the corresponding perforations. For the elementary mechanisms, the analytical ratio $M_{33}/M_{11}$ depends on the medium stiffness at the location of the seismic source (see Figure \ref{fig:vp}), which explains its variability across the perforation shots. The comparison with the ratio of the inverted MTs is only semi-quantitative. In fact, we do not account for the seismic wavelength effect when deriving the analytical ratio, which would have smoothed out the spatial variability of this ratio.

From Figure \ref{fig:mt_ratio}, it seems that source mechanism variability occurs throughout the reservoir layer since sequential shot numbering corresponds to consecutive source locations in Figure \ref{fig:vp}. To understand the spatial distribution of this variability within each stage, we show in Figure \ref{fig:mt_ratio_per_stage} the same ratio $M_{33}/M_{11}$ as function of stage number. The color indicates the sequential shot numbering within each stage (1 to 6 at most). Note that some stages are missing (no data available). We observe that low (high) $M_{33}/M_{11}$ ratio sources systematically correspond to early (later) perforations. This means that the source mechanism variability is unlikely to be caused by the pre-existing local reservoir heterogeneity (before stimulation) which should appear random across stages. Instead, the variability is tightly linked to the perforation/stimulation procedure. One possible conjecture is that early perforations in a given stage are the closest to the previously perforated and stimulated stage (Figure \ref{fig:sketch_perforation}). Thus, they are more effective in creating cracks in the part of the rock that is closest to the previous stage.

\plot[!h]{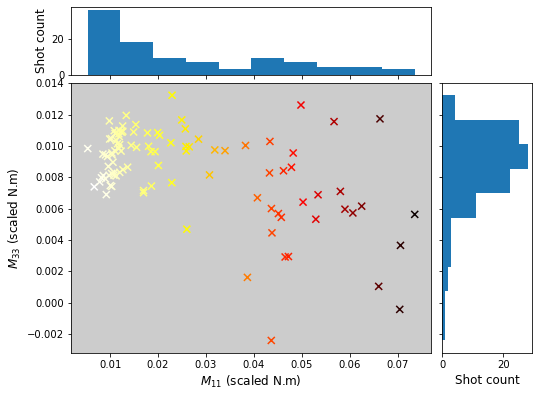}{width=0.95\textwidth}{Cross-plot of the inverted MT components with their respective histograms. Dark (bright) colors correspond to large (small) observable seismic moment. Two main perforation clusters can be observed with two different distributions of the MT components, suggesting a non-random change in source mechanism.}

\plot[!h]{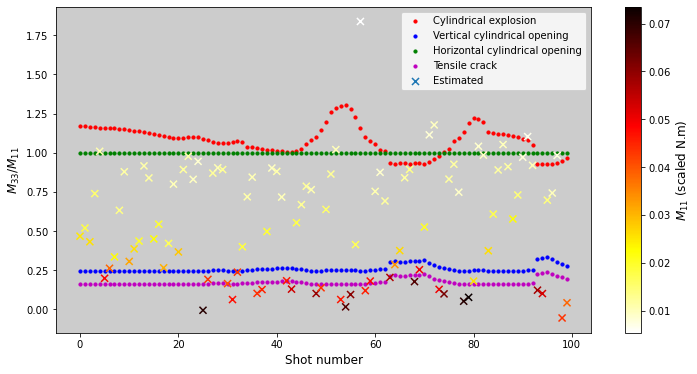}{width=0.95\textwidth}{Ratio $M_{33}/M_{11}$ for the inverted MTs and for four fundamental source mechanisms. High ratios indicate explosion-dominated sources whereas low ratios indicate a larger crack-opening mechanism. The value of $M_{11}$ in the latter case could serve as a relative measure of the crack width created by the perforations.}

\plot[!h]{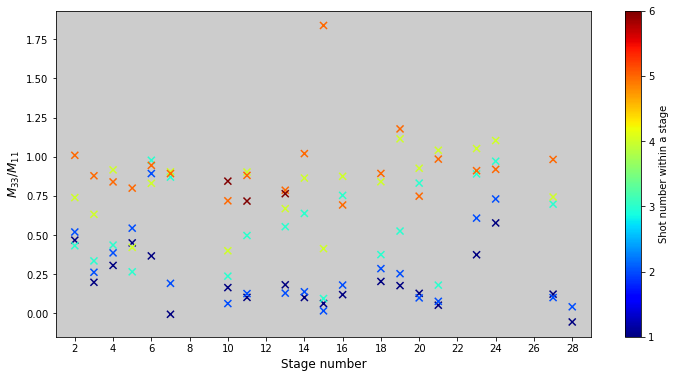}{width=0.95\textwidth}{Ratio $M_{33}/M_{11}$ for the inverted MTs as a function of stage number. Low and high ratios correspond systematically to early and late perforations in each stage. This indicates that source mechanism variability is tightly linked to the perforation/stimulation procedure.}
\section*{Conclusions}
Perforation shots in unconventional reservoir wells generate elastic waves that a downhole DAS fiber can record. We proposed a superposition of three mechanisms to represent such seismic sources and derived their corresponding moment tensors. We developed a workflow to estimate the resolvable components of the total moment tensor (MT) using DAS data. The vertical heterogeneity of the propagating elastic medium allows for resolving the $M_{33}$ component which cannot otherwise be resolved using a DAS fiber aligned with the perforations. A careful analysis of the inverted results shows that the MTs can indicate which perforations are more effective in creating micro-cracks in the surrounding rock. Moreover, the MTs exhibit a geometrical periodic pattern distinguishing early and later perforations in each stimulation stage. Our MT inversion yields a good fit with field data and can serve as a basis for subsequent elastic full-waveform inversion.
\append[appendix:mt]{Elementary moment tensors}

To derive the general expression of the elementary moment tensors representing the four source mechanisms described in Figure \ref{fig:sketch_mechanism}, we evaluate the integral \eqref{mt-general} over the appropriate dislocation surface $S$ for each component $M_{kl}$. For the cylindrical explosion and cylindrical opening, $S$ is the outer surface of a cylinder of some unknown characteristic length and the displacement jump $[u_i]$ is radial. For the dipole force, the MT is a simple force couple (tensile dipole) and $S$ reduces to two opposite points. For the tensile crack, $S$ is an arbitrary planar surface orthogonal to the $x$-axis and the displacement jump $[u_i]$ occurs in the $x$-direction. In all cases, we assume the displacement jump and the medium properties constant over the surface $S$. Thus, the MT expressions for a surrounding isotropic medium read
\begin{equation}\label{mt-individual-isotropic}
\begin{split}
{\mathbf{M}}^{ce} =M_0^{ce} \sqrt{2} \frac{\mathbf{T}^{ce}}{\Vert \mathbf{T}^{ce} \Vert_F} \quad &; \quad \mathbf{T}^{ce} =
\begin{pmatrix}
\lambda & 0& 0\\
 0 & \lambda + \mu &  0\\
0& 0 & \lambda + \mu 
\end{pmatrix},\\
{\mathbf{M}}^{df}(\theta)=M_0^{df} \sqrt{2} \mathbf{T}^{df} \quad &; \quad \mathbf{T}^{df} =
\begin{pmatrix}
0 &0 & 0\\
 0 & s^2 & -sc \\
0& -sc & c^2 
\end{pmatrix},\\
{\mathbf{M}}^{co}(\theta)=M_0^{co} \sqrt{2} \frac{\mathbf{T}^{co}}{\Vert \mathbf{T}^{co} \Vert_F} \quad &; \quad \mathbf{T}^{co} =
\begin{pmatrix}
\lambda + \mu & 0&0 \\
  0& \lambda + c^2\mu  & sc\mu \\
0& sc\mu & \lambda + s^2\mu
\end{pmatrix},\\
{\mathbf{M}}^{tc}=M_0^{tc} \sqrt{2} \frac{\mathbf{T}^{tc}}{\Vert \mathbf{T}^{tc} \Vert_F} \quad &; \quad \mathbf{T}^{tc} =
\begin{pmatrix}
\lambda + 2\mu & 0&0 \\
  0& \lambda & 0 \\
0& 0 & \lambda
\end{pmatrix},
\end{split}
\end{equation}
where $\lambda$ and $\mu$ are the Lam\'e parameters, $s=\sin(\theta)$, $c=\cos(\theta)$, $\theta$ is the phasing angle in the $y-z$ plane, and $M_0^{\dots}$ are the scalar seismic moments defined in expression \eqref{seismic-moment} and corresponding to each MT. $M_0^{\dots}$ depend on the amount of dislocation (displacement jump and surface area of $S$) caused by each source mechanism and are unknown \textit{a priori}. When multiple charges with different phasing angles $\theta_k$ are released simultaneously, we need to sum their individual MTs to obtain the expressions for $\mathbf{M}^{df}$ and $\mathbf{M}^{co}$, assuming there are no non-linear interactions between the charges. If the charges have the same power and cause similar dislocation, the resultant MTs will have the same expressions as in \ref{mt-individual-isotropic} after setting $c^2=\sum_k\cos^2(\theta_k)$, $s^2=\sum_k\sin^2(\theta_k)$, and $sc=\sum_k\sin(\theta_k)\cos(\theta_k)$ (summation over different charges).

For a surrounding VTI medium, the tensors $\mathbf{T}^{\dots}$ appearing in the expressions above become
\begin{equation}\label{mt-individual-vti}
\begin{split}
\mathbf{T}^{ce} &=
\begin{pmatrix}
C_{11}-2C_{66}+C_{13} & 0& 0\\
  0& C_{11}+C_{13} & 0 \\
0& 0 & C_{33}+C_{13} 
\end{pmatrix},\\
\mathbf{T}^{co}(\theta)&=
\begin{pmatrix}
(C_{11}-C_{66}) &0 &0 \\
 0 & c^2(C_{11}-C_{66}) + s^2C_{13} & sc(C_{11}-C_{66}-C_{13}) \\
0& sc(C_{11}-C_{66}-C_{13}) & s^2(C_{11}-C_{66})+c^2C_{13}
\end{pmatrix},\\
\mathbf{T}^{tc}&=
\begin{pmatrix}
C_{11} &0 &0 \\
  0& C_{12} & 0 \\
0&0  & C_{13}
\end{pmatrix},
\end{split}
\end{equation}
where $C_{ij}$ are the elements of the stiffness tensor in Voigt notation. The tensor $\mathbf{T}^{df}$ remains unchanged.

If the angle $\theta$ is known, the number of unknowns in the total MT \eqref{mt-total} reduces to three which correspond to the scalar seismic moments $M_0^{ce}$, $M_0^{df}$, and $M_0^{co}$. However, it will still be not possible to resolve those three parameters using a DAS fiber aligned with the perforation shots in the same horizontal well.

\bibliographystyle{seg}


\end{document}